\documentclass[10pt]{iopart}
\usepackage{epsfig}
\epsfclipon

\begin{document}

\title{Kaon and $\Phi$ production vs Participants in Nuclear Collisions}

\author{H Bia\l kowska\dag\ and W Retyk\ddag}
\address{\dag\ Institute for Nuclear Studies, Warsaw, Poland}
\address{\ddag\ Institute of Experimental Physics, Warsaw University, Warsaw, Poland}

\begin{abstract}
Data on kaon and $\Phi$ production in nuclear collisions as a function of 
centrality are analysed both at AGS and SPS energy range.  
We compare the results of several experiments, looking for
common trend in `participant scaling' of production yields. We find a smooth
description of scaled kaon and $\Phi$ yields as a function of participant 
density.  We also show a participant density dependence of kaons 
and $\Phi$ produced in the forward hemisphere for proton-nucleus collisions.
\end{abstract}

%\submitto{\jpg}

%\maketitle

There exists a large amount of data, both at AGS and SPS, where charged
kaon production in nuclear collisions is studied. Kaons represent between
70 and 80 \% of all strangeness produced, and thus carry important
information on global strangeness.

The main physics question we want to address is strangeness evolution
with centrality.

The idea of centrality is a geometrical one, and it refers to an impact
parameter b of the colliding nuclei. In real life, impact parameter is not
accessible to measurement. One has to measure some quantity x related to
b, use some model to translate x into b, and only then one can calculate
the number of participants in a collision, $N_{w}$.

Variable x can be the transverse energy $E_{T}$(as used by NA50), charged
particle multiplicity $N_{ch}$ (used by WA97) and forward, or zero degree
energy, $E_{veto}$, used by AGS experiments and NA49.

When studying global strangeness characteristics, it is important to look
at $4\pi$ multiplicity, and it often involves some extrapolation. In the
following we will show kaon data from AGS \cite{BNLPRC}, NA49 \cite{FerQM} 
and WA97 \cite{WA97QM}, \cite{Nicola}. AGS data and NA49 data are measured
and extrapolated by the respective authors in the full phase space. The
WA97 data are $p_{T}$ integrated but measured/extrapolated to only one
unit of rapidity, centered at midrapidity. In order to compare with NA49,
we will extrapolate these data to $4\pi$, assuming same scaling factor 
from midrapidity to $4\pi$ as in NA49.

\begin{figure}[htbp]
 \begin{center}
  \begin{tabular}{c}
   \begin{minipage}[b]{0.5\textwidth}
    \centerline{\epsfxsize=0.9\textwidth \epsfbox{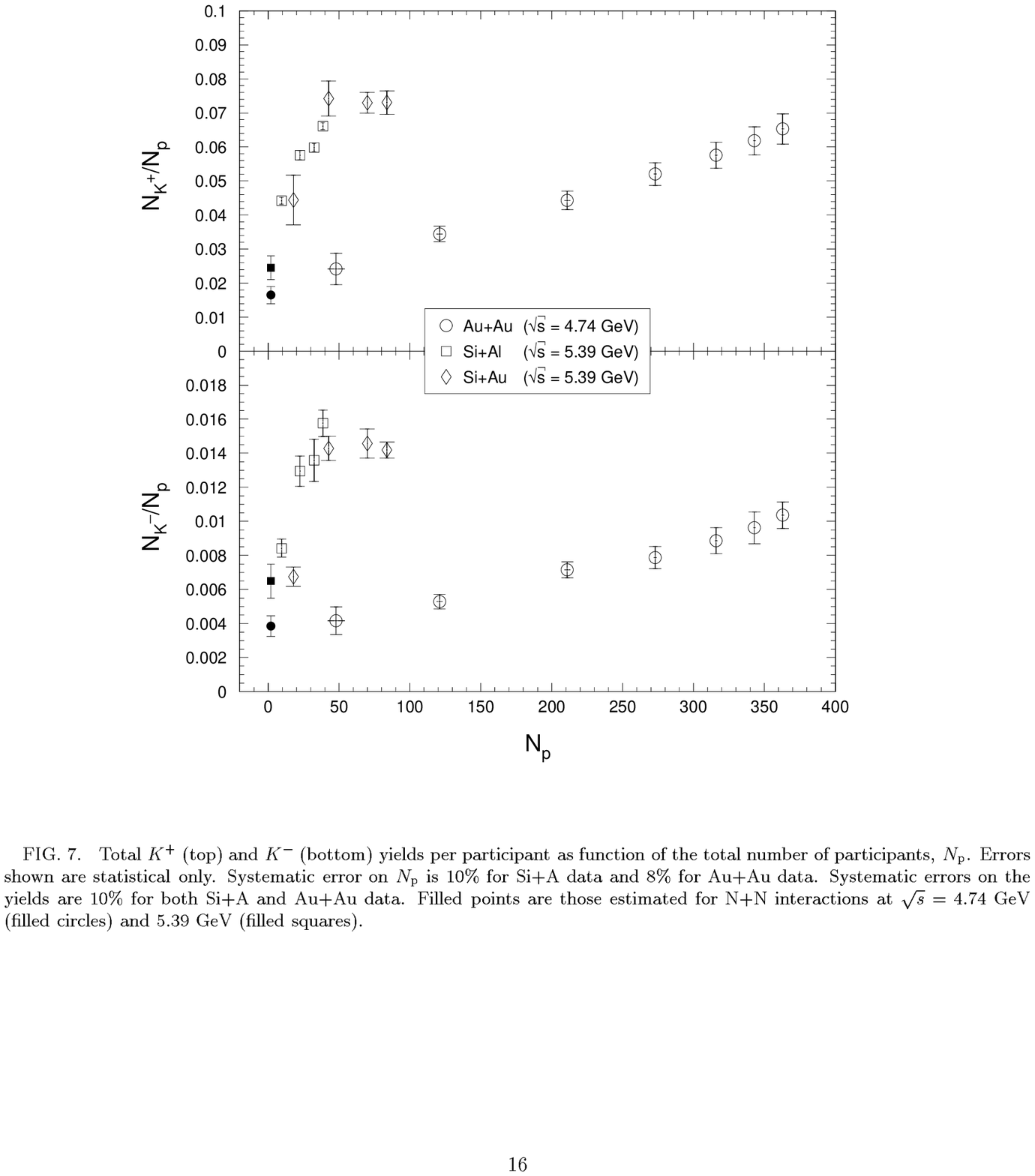}}
    \caption{Charged kaon per participant yields vs number of participants 
             at AGS energies \cite{BNLPRC}.}
    \label{fig:BNLPRC}
   \end{minipage}
   \begin{minipage}[b]{0.5\textwidth}
    \centerline{\epsfxsize=0.9\textwidth \epsfbox{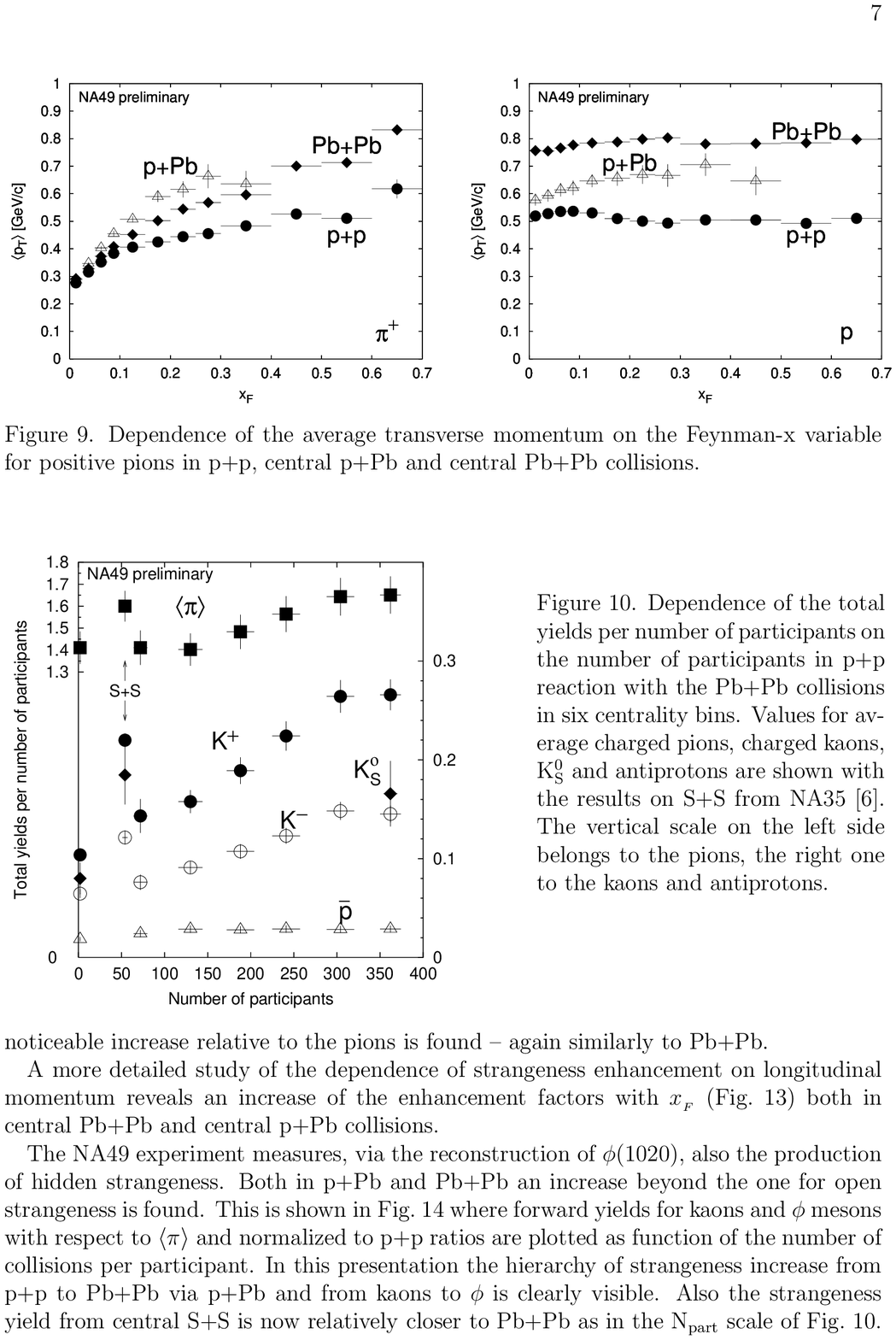}}
     \caption{Average of charged kaons, pions and anti-protons 
              per par\-ti\-ci\-pant vs number of par\-ti\-ci\-pants at SPS 
              energies (NA49) \cite{FerQM}. Points other than marked 'S+S' come
              from Pb+Pb reaction.}
     \label{fig:FerQM}
   \end{minipage}
  \end{tabular}
 \end{center}
\end{figure}

Fig.~\ref{fig:BNLPRC} \cite{BNLPRC} shows the charged kaon per participant 
yields vs participant number in Au~-~Au, Si~-~Al and Si~-~Au collisions at AGS
energies. The number of participants is determined directly from the total
spectator energy, measured by the forward calorimeter.

\epsfclipon
 
Fig.~\ref{fig:FerQM} \cite{FerQM} shows the average of charged kaons per
participant vs participant number from Pb~-~Pb collisions at 158 GeV/c.
Here the number of participants was
determined directly as a (measured) total number of net baryons, that is
baryons minus antibaryons. 

Both plots show the same tendency. The yield of charged kaons per
participant increases significantly with centrality, as measured by the
participant number. Both at AGS and SPS the number of participants does
not represent a good scaling variable. Apparently the same number of
participants does not represent the same geometry for different nuclei;
clearly, dilute surfaces of big nuclei make a difference.

H.Satz and M.Nardi \cite{Nardi} have suggested to use the average density
of participants in nuclear collisions as a centrality parameter. The
following formulae give total number of participants
and average density of participant nucleons at impact parameter b for
nucleus A~-~nucleus B collision.

\begin{eqnarray*}
\rho^{^{AB}}_{w}(\vec b,\vec s) & = &
 A  T_{\!_A}(\vec s)\! \left\{
1- \left[
1-\sigma_{\!_0}\, T_{\!_B}(\vec b -
\vec s)\right]^{\scriptscriptstyle\! B}\right\} \\
& + & B\; T_{\!_B}(\vec b-\vec s)
\left\{  1- \left[
1-\sigma_{\!_0}\, T_{\!_A}(
\vec s)\right]^{\scriptscriptstyle\! A}\right\}
\end{eqnarray*}

\[
\langle \rho^{^{AB}}_{w}(b) \rangle =
{\int \mathrm{d}^2s [\rho^{^{AB}}_{w}(\vec b,\vec s)]^2 }/
{\int \mathrm{d}^2s\; \rho^{^{AB}}_{w}(\vec b,\vec s)}~.
\]

\epsfclipon
\begin{figure}[htbp]
 \begin{center}
  \begin{tabular}{c}
   \begin{minipage}[b]{0.5\textwidth}
    \centerline{\epsfxsize=0.9\textwidth \epsfbox{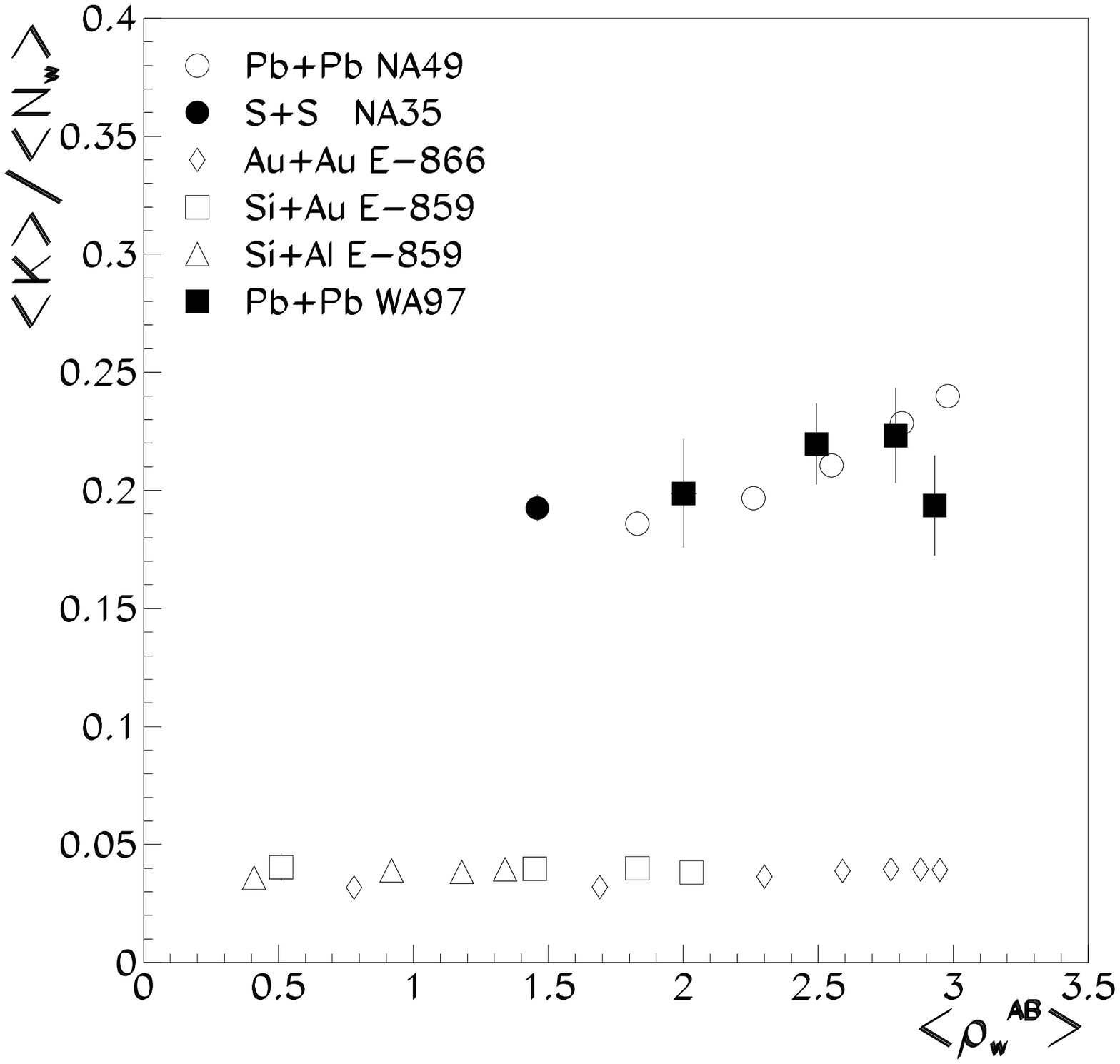}}
    \caption{The average yield of charged kaons per par\-ti\-ci\-pant vs 
             par\-ti\-ci\-pant density from differ\-ent experiments 
             (extrapolated to 4$\pi$).}
    \label{fig:wb5}
   \end{minipage}
   \begin{minipage}[b]{0.5\textwidth}
    \centerline{\epsfxsize=0.9\textwidth \epsfbox{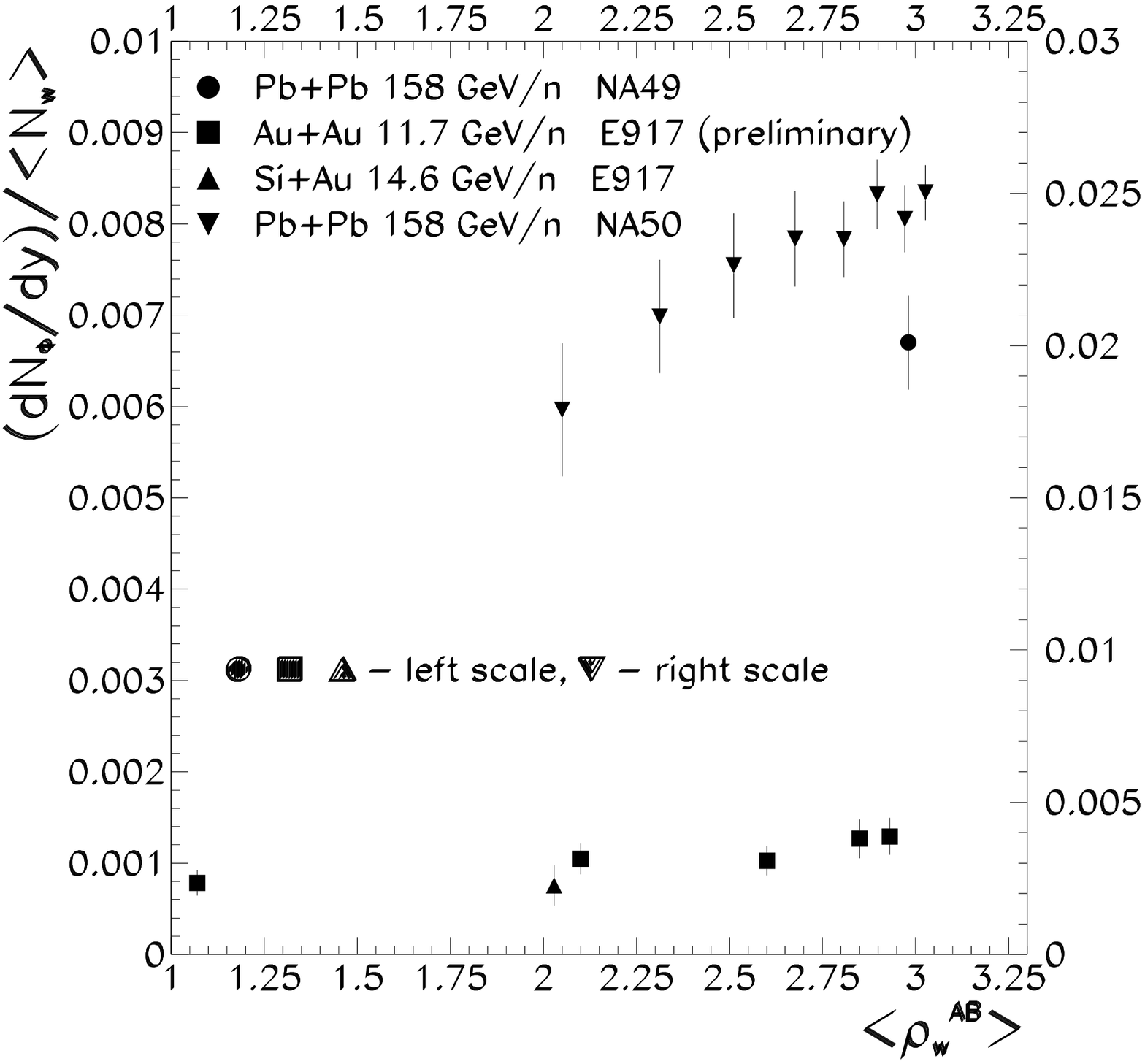}}
     \caption{$dN/dy$ of $\Phi$ mesons per par\-ti\-ci\-pant vs
             par\-ti\-ci\-pant density from different experiments. 
             Data sets has different scales.}
     \label{fig:wb13}
   \end{minipage}
  \end{tabular}
 \end{center}
\end{figure}

In Fig.~\ref{fig:wb5} we show the average yield of charged kaons per 
participant vs participant density for all data sets discussed. 
Here we have taken the
same definition of participants for all experiments, and have calculated
the average participant density on the basis of percentage of
$\sigma_{inelastic}$  and hence $b$, taking Glauber formulae with
Saxon~-~Woods density.
Notice that it changes slightly the number of participant used in all
quoted experiments, but assures equal treatment of all data sets.

Two observations are in order. First, the average participant density
seems to take out large part of difference between collisions of different
nuclei. Second, there still seems to be an increase of kaon/participant
with participant density at SPS, although most central WA97 point is
different from NA49.

These observations have encouraged us to look at the $\Phi$ meson
production as a function of centrality. $\Phi$ is much less copiously
produced, but it carries a potentially very important information on the 
production of $s\bar{s}$ state
 
At AGS the E917 experiment measures $\Phi$ at 11.7GeV/c/N. On This
Conference the authors have shown new (although still
preliminary) data, which they kindly have allowed us to use
\cite{E917new}. These data come from limited rapidity range,
$1.2<y<1.6$.

The NA50 experiment \cite{NA50QM} has shown the $\Phi$ production
dependence on the
number of participants in Pb~-~Pb collisions at 158 GeV/c. This experiment
measures the $\mu^{+}\mu^{-}$ decay channel, and covers midrapidity
range  and large $m_{T}$, bigger than \hbox{1.5 GeV/c}. 
Finally, NA49 has measured
$\Phi$ production in the $K^{+}K^{-}$ decay channel, with large rapidity
and transverse momentum coverage, in central Pb~-~Pb ( 5\% of
$\sigma_{inel}$), \cite{PhiPLet}.

We have recalculated the $\Phi$ yields, integrating the transverse
spectra
and selecting (approximately) the same scaled rapidity range. The average
participant density and number of participants were  also calculated
starting from b and
$N_{w}$ range. Fig.~\ref{fig:wb13} shows the compilation of all available 
AGS and SPS data.
 
  An increase of $\Phi$ per participant yields with the
average participant density is seen at AGS
energy, and a significant increase is observed from SPS data.
For the highest centrality Pb~-~Pb data, there is a discrepancy in the
absolute $\Phi$ yield - NA50 results exceed NA49 yield by factors of 
between 3 and 4.

We would like to extend this study into proton-nucleus collisions.
For these reactions, measurement of slow (`grey' in old emulsion
terminology) protons serves as a centrality trigger. The measured number
of slow protons can be translated into the number of collisions $\nu$,
assuming independence of successive collisions. The full number of
participants is then $1+\nu$. 
Using the same definition as for nucleus-nucleus collisions, 
participants density becames in the case of proton-nucleus reaction the above 
defined number of participants.

For the proton~-~nucleus collisions, there is an important difference
between center of mass hemispheres. In the backward hemisphere effects of
target fragmentation, energy loss and cascading contribute to the overall
multiplicity. In the forward hemisphere we deal with fragmentation of
projectile proton that has hit several nucleons. At midrapidity we
have contributions from both mechanisms, or a `spill~-~over'.
 
AGS experiment E910 (\cite{E910prep}, \cite{E910new}) has measured $\Lambda$
and $K^{0}$ production for
centrality selected p~-~Au collisions at 17.5 GeV/c. They observe a
significant rise of strange particle multiplicity with the number of
collisions, $\nu$, saturating for higher $\nu$.

\begin{figure}[htbp]
 \begin{center}
  \begin{tabular}{c}
   \begin{minipage}[b]{0.8\textwidth}
    \centerline{\epsfxsize=0.9\textwidth \epsfbox{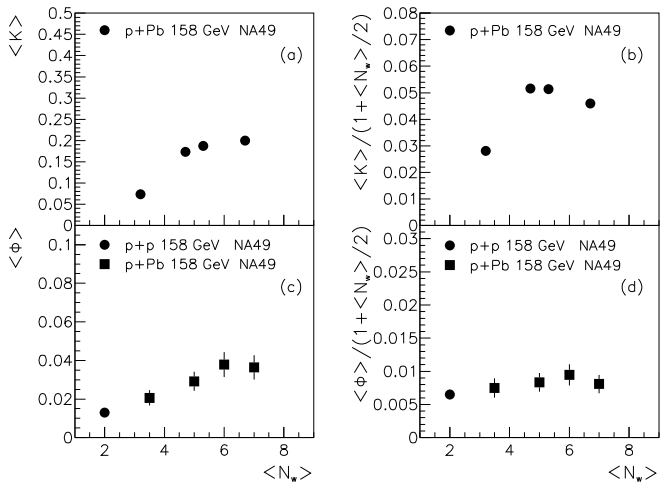}}
    \caption{The average multiplicity of charged kaons (a,b) and 
             $\Phi$ mesons (c,d) in the forward hemisphere 
             as a function of the number of participants. In the right 
             panels multiplicity is normalized in order to account for the 
             influence of the target fragmentation as a function of the 
             number of participants.}
    \label{fig:wbz}
   \end{minipage}
  \end{tabular}
 \end{center}
\end{figure}

SPS experiment NA49 has measured charged kaon production in centrality
selected p~-~Pb collisions at 158 GeV/c \cite{Hoehne} and \cite{PhiPLet}.
Here the rapidity coverage extends over forward hemisphere only. Thus we
deal both with the projectile fragmentation and a spill~-~over from the
target hemisphere. Fig.~\ref{fig:wbz}a shows the average charged kaon 
multiplicity as a function of the number of participants $N_{w}$. 
In order to account
for the influence of the target fragmentation, we propose to normalize the
kaon yield not by $N_{w}$ but by the $N_{w}/2+1$, as shown in
Fig.~\ref{fig:wbz}b.
We observe a definite kaon enhancement. 

Experiment NA49 has also measured $\Phi$ (in the $K^{+}K^{-}$
decay channel) production in centrality selected p~-~Pb collisions 
\cite{PhiPLet}.
Fig.~\ref{fig:wbz}c shows the average yield of $\Phi$ (in the forward 
hemisphere) as a
function of the number of participants, and Fig.~\ref{fig:wbz}d - $\Phi$ yield
normalized by $(N_{w}/2+1)$. A slow increase from pp point, measured
in the same experiment, with $N_{w}$ is seen.

\begin{figure}[htbp]
 \begin{center}
  \begin{tabular}{c}
   \begin{minipage}[b]{0.5\textwidth}
    \centerline{\epsfxsize=0.8\textwidth \epsfbox{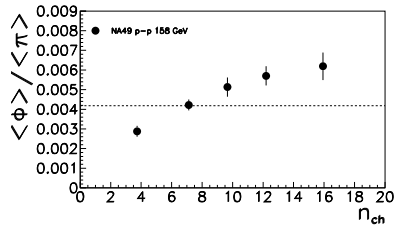}}
   \end{minipage}
   \begin{minipage}[b]{0.5\textwidth}
    \caption{The average $\Phi$ yield normalized to the average pion yield
             in proton~-~proton collisions at \hbox{158 GeV/c} as a function
             of charged particle multiplicity. The horizontal dashed line
             indicates the cross-section weighted average.}
    \label{fig:phi2pi}
  \end{minipage}
  \end{tabular}
 \end{center}
\end{figure}

As a last observation we would like to show a completely new approach to
$\Phi$ enhancement. Fig.~\ref{fig:phi2pi} shows the average $\Phi$ 
yield normalized to the average pion yield in
proton~-~proton collisions at \hbox{158 GeV/c}, studied as a function 
of charged particle multiplicity \cite{PhiPLet}. 
For these data there is no asymmetry of the CM system, and
measurement in the forward hemisphere allows for conclusions for the full
phase space. While it is hard to define `centrality' for proton~-~proton
collisions, we find it interesting to observe `strangeness enhancement' as
a function of - presumably - `violence' of the pp collision.

To sum up, we make the following observations:
\begin{itemize}
\item Participant density allows for smooth comparison of various nuclear
collisions
\item At AGS average kaon yield per participant stays approximately
constant as a function of participant density
\item At SPS there is a hint for an increase of kaon/participant as a
function of participant density
\item The $\Phi$ yield per participant definitely increases with
participant density both at AGS and SPS
\item Both kaon and $\Phi$ yields increase with participant density in
p~-~nucleus collisions
\item $\Phi$ production in proton~-~proton collisions normalized to the 
average pion yield increases with charged particle multiplicity
\end{itemize}
The above observations have to be taken into account when attempting to
describe and interpret strangeness enhancement in nuclear collisions.

\ack
We would like to thank The Organizers of this Conference and in particular
its Chairperson for excellent organization and truly scientific
atmosphere. HB thanks for financial support at the Conference.  
We thank our colleagues from the NA49 experiment for the permission to use
new results on $\Phi$ production.
This work has been supported by the Polish State Committee for Scientific
Research \hbox{(2 P03B 099 16)}.

%\section{Figure Captions}
\end{document}